\documentclass[showpacs,superscriptaddress,floatfix,twocolumn,prl]{revtex4}

\usepackage{amsmath}
\usepackage[dvips]{graphicx}
\usepackage{epsfig}
\usepackage{tabularx}

\begin{document}

\title{Coexisting attractors and chaotic canard explosions in a slow-fast optomechanical system}
\author{Francesco~Marino}
\affiliation{CNR-Consiglio Nazionale delle Ricerche, Istituto dei Sistemi
Complessi, via Madonna del Piano 10, I-50019 Sesto Fiorentino,
Italy}

\author{Francesco~Marin}
\affiliation{Dipartimento di Fisica e Astronomia, Universit\`a di Firenze, INFN
Sezione di Firenze, and LENS,\\ Via Sansone 1, I-50019 Sesto Fiorentino
(FI), Italy}

\date{\today}
\begin{abstract}
The multiple timescale dynamics induced by radiation pressure and photothermal effects in a high-finesse optomechanical resonator
is experimentally studied. At difference with 2D slow-fast systems, the transition from the
quasi-harmonic to the relaxational regime occurs via chaotic canard explosions, where large-
amplitude relaxation spikes are separated by an irregular number of subthreshold oscillations.
We also show that this regime coexists with other periodic attractors, on which the trajectories evolve on a substantially faster time-scale.
The experimental results are reproduced and analyzed by means of a detailed physical model of our system.

\end{abstract}

\pacs{05.45.-a, 42.65.Sf, 42.65.-k}

\maketitle

\section{introduction}

Slow-fast systems, i.e. nonlinear dynamical systems in which two or more processes are governed by very different time scales,
appear in many branches of natural science \cite{multiple}. Typically two-dimensional (2D) systems undergo a supercritical
Hopf bifurcation in which the attractor changes from a stable equilibrium to
stable relaxation oscillations. Two kinds of oscillatory regimes can be readily identified, as paradigmatically illustrated
by the FitzHugh-Nagumo equations \cite{fhn}. In a narrow parameter range beyond the Hopf bifurcation,
a small-amplitude quasi-harmonic limit cycle arises (subthreshold oscillations),
with frequency given by the imaginary part of the linear eigenvalues of the fixed point.
Out of this range, the amplitude of the limit cycle abruptly-- though continuously-- change
to that of the relaxation oscillations regime, with a frequency depending on the splitting between the time scales.
Here, the dynamics can be decomposed into a sequence of periods of slow motion,
taking place near the attracting branches of the slow-manifold (equilibria of the fast subsystem) \cite{smale},
separated by faster relaxation jumps between them.
The transition from the quasi-harmonic to the relaxational regime, referred to as canard explosion \cite{canard},
occurs within an exponentially small range of a control parameter, in which the system trajectories closely follow
for some time also the repelling part of the slow manifold.

In high-dimensional phase spaces, canard explosions give rise to more complex scenarios \cite{kramer,desroches}.
While 2D systems display subthreshold or relaxation oscillations but never both of them for a fixed set of parameters,
in higher dimensions a clear separation of the two regimes is usually lost. This is the case of
mixed-mode oscillations \cite{mmo}, consisting of an alternation of relaxation excursions and (small-amplitude) quasi-harmonic oscillations.
These regimes can be periodic or chaotic, in which case relaxation spikes are separated by
an irregular number of subthreshold oscillations (chaotic spiking).
Similar spiking sequences have been observed in a semiconductor laser
with optoelectronic feedback \cite{kais1,epjd} and, recently, in a high-finesse suspended-mirror optical cavity \cite{prenoi}.

In the latter, the intracavity field and the mirror motion are coupled through radiation pressure and photothermal effect (the latter indicating the
thermal expansion of the mirrors due to the absorbed intracavity light). Since such processes, as well as the intracavity field itself,
are competing and are governed by very distinct time-scales, optomechanical resonators possess all the typical features of slow-fast dynamical systems.
Previous experiments in optomechanical resonators have shown that periodic and chaotic oscillations
can be induced by either radiation-pressure \cite{in6,in2,in7} or photothermal nonlinearities \cite{in8}. In all these cases,
regenerative oscillations arise thanks to the delay of the electric field buildup in the cavity or due to the
spatio-temporal profiles of the coupled elastic and temperature fields \cite{in10}.
Recently, also the interplay between radiation-pressure and changes to mechanical frequency due to absorption
heating has been considered \cite{in9}.
However, the regime in which both radiation pressure and photothermal effects are relevant has been poorly explored so far.
The occurrence of irregular spike sequences induced by these effects has been theoretically predicted in Ref. \cite{noi}, in a regime where
the mirror motion was overdamped and the field adiabatically eliminated. In Ref. \cite{prenoi}, we provided experimental evidence of these phenomena, although in a completely different regime of parameters. However, important features of the chaotic spiking regime,
in particular its relation to canard explosions in a high-dimensional phase-space, were not exhaustively analyzed.

Here we extend and generalize the results of Ref. \cite{prenoi}, unveiling the effects of the multiple timescale competition between photothermal and radiation pressure nonlinearities in an optomechanical resonator. In particular, we show that chaotic spiking sequences
are the result of canard explosions in a higher-dimensional phase space, in which trajectories are rapidly attracted towards the stable branches of
a one-dimensional {\bf S}-shaped manifold, where the slow dynamics takes place. We also show that this regime coexists with other non-trivial attractors,
on which the trajectories evolve on a substantially faster time-scale. Therefore, which attractor is eventually reached by the system depends only on
the choice of the initial conditions. Finally, the experimental results are reproduced by means of a detailed physical model,
which allows us to identify the mechanims underlying the observed dynamics.
Since the stable operation of optomechanical resonators is a fundamental requirement in many quantum optics experiments \cite{qgs1,qgs2},
a complete characterization of the system attractors is mandatory, in order to identify the instability boundaries as well as
to design suitable control schemes.

\section{Experimental Setup}

\begin{figure}
\begin{center}
\includegraphics*[width=1.0\columnwidth]{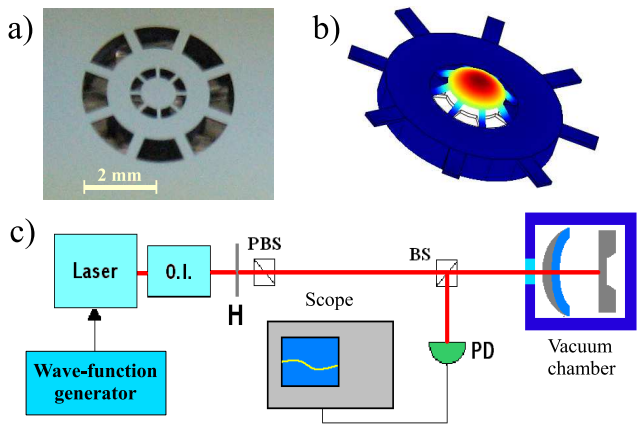}
\end{center}
\caption{a) Image of the double wheel oscillator. b) FEM calculation of the fundamental mode. c) Basic scheme of the experimental apparatus. O.I.: optical isolator; H: half-wave plate; PD: photodiode; PBS: polarizing beam-splitter; BS: beam-splitter.}
\vspace{-.3cm}
\label{figu1}
\end{figure}

Our oscillating mirror, realized on a 500~$\mu$m silicon substrate, consists of a double wheel (see Fig. \ref{figu1}a,b).
The central mass (main oscillator) and the arms are 70~$\mu$m thick, while the intermediate ring mass (isolating stage)
is 500~$\mu$m thick. These particular structures allow us to reduce the mechanical coupling between the oscillator fundamental mode (see Fig. \ref{figu1}b) and high-frequency modes of the substrate \cite{japserra}. On the front side of the wafer, a deposition of alternate Ta$_2$O$_5$/SiO$_2$ quarter-wave layers provides the highly reflective coating.
The main oscillator is used as end mirror of a $L=12.2$~mm long Fabry-Perot cavity
with a $50$~mm radius silica input mirror (transmissivity $\mathcal{T}$=110 ppm)
operating in a vacuum chamber at $5\cdot10^{-4}$~Pa. We obtain a cavity finesse of $\mathcal{F}$ = $3\cdot10^{4}$ (half linewidth $\gamma = 225$~kHz), with an
intracavity optical power $P_c=1.1\cdot10^{4}P_{in}$, where $P_{in}$ is the optical power at the input mirror.
The micro-mirror has a mechanical frequency $\omega_M / 2 \pi \simeq $~250~kHz, a mechanical quality factor
$Q \simeq$~5~$10^{3}$ and an effective mass $m \simeq ~10^{-7}$~Kg.

The aim of the experiment is to explore the system dynamics as the detuning between laser and cavity
resonance, $\Delta \nu$, and the injected optical power are varied.
To this end, the laser is free-running in order to avoid possible modifications of the dynamics
induced by the frequency locking servo-loop.
The experimental setup is sketched in Fig.~\ref{figu1}c.
The light source is a cw tunable Nd:YAG laser operating at $\lambda$=1064~nm.
After a 40~dB optical isolator, the beam is injected into the cavity.
The radiation intensity is controlled by an half-wave plate followed by
a polarizing beam splitter and the reflected signal beam is monitored by the photodiode PD.
The laser frequency can be continuously tuned within a range of 6 GHz by changing the temperature of the monolithic laser crystal through an offset voltage.
Fast frequency scanning is achieved by applying a ramp voltage signal to a PZT on the laser crystal.

\section{Theoretical model}

In an optomechanical resonator, the intracavity field
and the cavity length variations are nonlinearly coupled through radiation pressure and photothermal effects.
A physical model of an optomechanical resonator in the presence of these two effects has been derived in Ref. \cite{noi1} and
subsequently analyzed in Ref \cite{noi2}. The radiation pressure, pushing the mirrors, tends to increase the cavity length with respect to
its cold-cavity value (i.e. in absence of radiation pressure and photothermal effects).
Therefore, by injecting a monochromatic beam detuned on the long
wavelength side with respect to the cold-cavity resonance, the cavity resonance moves towards
the injected wavelength and therefore the intracavity intensity increases.
The intracavity field has the further effect of varying the temperature of the mirrors. Heating induces a decrease
of the cavity length through the mirror thermal expansion, and it moves away the cavity resonance
thereby decreasing the intracavity intensity. It is worth to notice that radiation-pressure acts on a fast timescale,
given by $2 \pi /\omega_M$ (about 4 $\mu$s in our resonator), while the photothermal effect on the much slower $2 \pi /\omega_c \approx 20$ms, 
where $\omega_c$ is the photothermal response angular frequency \cite{Braginsky}. 

The structure and the stability of the slow manifold is analogous
to that of the FitzHugh-Nagumo equations, thus giving rise to excitability and canard-orbits \cite{noi1}.
A radically new scenario for the transition from excitability to relaxation oscillations is observed
when the mirror motion is not overdamped, i.e. when the damping rate $\omega_M / Q$ becomes sufficiently smaller than $\omega_M$.
Now the fast motion includes inertial terms, therefore occurring on a two-dimensional fast manifold and the whole dynamics is
governed by a 3D slow-fast system. The initial Hopf bifurcation is followed by a cascade of period doubling bifurcations producing a sequence of
small periodic and chaotic attractors, that develops before relaxation
oscillations arise. As the mean amplitude of the chaotic attractors grows, a canard regime sets in, where
the small chaotic background spontaneously triggers excitable spikes in an erratic but deterministic sequence.
Such a regime has been theoretically studied in Ref. \cite{noi}. However, at difference with the model analyzed there,
in the system explored in this work the optical field evolves on a timescale comparable to that of
radiation pressure and cannot be adiabatically eliminated. The model thus requires to include also the intracavity field equation and reads
\begin{eqnarray}
\dot{\varepsilon} = -[1 - i (\delta_0 + \phi + \theta)] \varepsilon + 1 \; , \label{eq1} \\
\ddot{\phi} + \frac{\omega}{Q} \dot{\phi} + \omega^2 \phi = \alpha \vert \varepsilon \vert^2 \; , \label{eq2} \\
\dot{\theta} = -\omega_{PH} (\theta + \beta \vert \varepsilon \vert^2)  \; , \label{eq3}
\end{eqnarray}
where $\varepsilon$ is the intracavity field normalized to its resonant value; $\phi$ and $\theta$ are the cavity length changes due to radiation-pressure and photothermal expansion, respectively, normalized to $\gamma_l=\gamma \frac{\lambda L}{c}$;
$\delta_0$ is the detuning between laser and cavity resonance for zero input power, normalized to $\gamma$;
$\omega = \omega_M / \gamma$ and $\omega_{PH} = \omega_c / \gamma$ are the dimensionless oscillator and photothermal frequencies.
The parameters $\alpha = \frac{2 P_c}{m c \Gamma^2 \gamma_l}$ and
$\beta$ measure the strength of the radiation pressure and photothermal effects \cite{noi1,noi2}.
Using the opto-mechanical parameters given before, we calculate $\omega=1.2$ and $\alpha=4.9~10^{-2}P_{in}$, with $P_{in}$ expressed in mW.
Concerning the photothermal effect, here we remark that Eq. (\ref{eq3}) realistically describes the photothermal dynamics
only at frequencies well above $\omega_{PH}$. Indeed, the frequency response of the length variations to a
modulated intracavity power has a slow, logarithmic divergence at low frequencies ~\cite{cerdonio,marin},
while it follows an intuitive 1/frequency-dependence as in a standard single-pole system at high frequencies.
The photothermal response is dominated by the fused silica input mirror for which we calculate $\omega_c/ 2 \pi \sim$~50~Hz,
leading to $\omega_{PH}=2.4~10^{-4}$, and $\beta=3.5~10^{-1}P_{in}$, given an absorption of 2 ppm.

\section{Experimental and Numerical results}
\begin{figure}
\begin{center}
\includegraphics*[width=1.0\columnwidth]{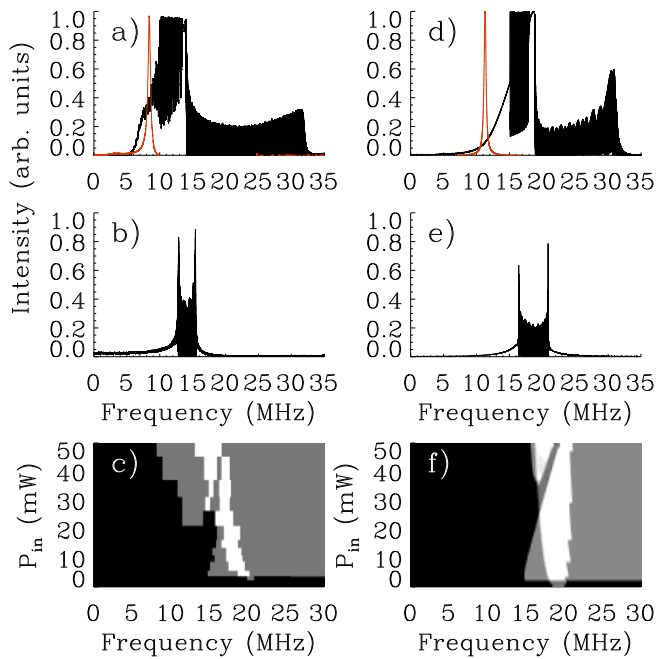}
\end{center}
\caption{Normalized reflected signal as the laser frequency is a) increased and b) decreased (scanning rate 6 MHz/ms) for $P_{in}$= 40 mW. Red (grey) trace
in a) show the Lorentzian reflection dip obtained for $P_{in}$=0.1 mW. c) State diagram reconstructed from reflected signal spectra as $P_{in}$ is varied:
In black regions the system has one stable fixed point. Grey regions refer to the coexistence of a fixed point and a non-stationary state
and white regions to the coexistence of oscillating solutions. Normalized reflected intensity $I = \vert \varepsilon \vert^2$ as $\delta_0$ is linearly d) increased and e) decreased in time
(rate of 6 MHz/ms), during the numerical integration of Eqs. (\ref{eq1})-(\ref{eq3}). Parameters: $\omega=1.2$, Q=5000, $\alpha=4.9~10^{-2}P_{in}$,
$\omega_{PH}=2.4~10^{-4}$ and $\beta=3.5~10^{-1}P_{in}$ with $P_{in}$=40 mW. Red (grey) trace in d) show the Lorentzian reflection dip obtained for $P_{in}$=0.1 mW.
f) State diagram obtained from numerical reflection spectra.}
%\vspace{-.3cm}
\label{figu2}
\end{figure}

We first characterize the stable and unstable regions of the system, as both the
laser frequency and injected power, $P_{in}$, are varied. Since the laser is not frequency-locked, the noise present in the system prevents the
point-by point construction of a bifurcation diagram. However, information about the instability domains can be roughly extracted from the system response
to a slow modulation of the laser frequency, by keeping fixed the injected power.
In Fig. \ref{figu2}a we show the reflected intensity as the laser frequency is swept,
starting from the red side (low frequencies) of the cavity resonance. For simplicity, the intensity signals are normalized in order to be zero
out of resonance and equal to 1 at resonance. For low input power we observe the typical Lorentzian reflection dip (red trace in Fig. \ref{figu2}a).
For high powers, instead, the line is strongly asymmetric and dynamical instabilities arise at both sides of the cavity resonance
(black trace in Fig. \ref{figu2}a). Around resonance, where the normalized optical intensity is about 1, the system is stable.
A completely different situation is found when the laser frequency is swept in the opposite direction (Fig. \ref{figu2}b).
Now, the width of the instability region is narrower, but a stable regime around resonance is never reached.
This is evidenced by the mean value of the optical spectrum, which remains much smaller than 1 in the whole
frequency range.

A complete stability diagram, reconstructed from the reflected intensity curves corresponding to different values of $P_{in}$, is
displayed in Fig. \ref{figu2}c). Such reconstruction does not allow a precise identification of the bifurcation boundaries, since
they might be dependent on the scanning rate, and narrow stability (or instability) domains could remain unresolved.
However, it allows us to determine whether the system possesses coexisting attractors and to roughly estimate their position
in the parameter space. In black regions the system is stable. Grey regions refer to the coexistence of a fixed point
and a non-trivial state, i.e. a limit cycle or a chaotic attractor. In white regions there are no stable fixed points and the
bistability involves oscillating regimes. In these situations, which attractor is eventually reached by the system trajectories
depends only on the choice of the initial conditions. In particular, notice that, depending on the injected power, the steady
resonant condition can coexist with an oscillating solution. As evidenced by optical spectra in Fig. \ref{figu2} a),b),
the stable operation at the cavity resonance is reached only by approching it from the red-side of the spectrum.

Fig. \ref{figu2} d),e) shows the reflected intensity signals as obtained by numerical integration of Eqs. (\ref{eq1})-(\ref{eq3})
with $\delta_0$ linearly varying in time. As in the experiment, for high powers and incresing $\delta_0$, the line shape becomes
strongly asymmetric with dynamical  instabilities at both sides of a narrow stability region around resonance.
On the other hand, as $\delta_0$ is decresed, the stable resonant condition is not found. The optical spectra and the stability
diagram in Fig. \ref{figu2} f), show that the line shapes and the widths of the regions of coexistence slightly differ from those obtained in the experiment.
This is probably due to the single-pole description of the photothermal effect that, for such slow scanning rates, is no longer a good approximation.

\begin{figure}
\begin{center}
\includegraphics*[width=1.0\columnwidth]{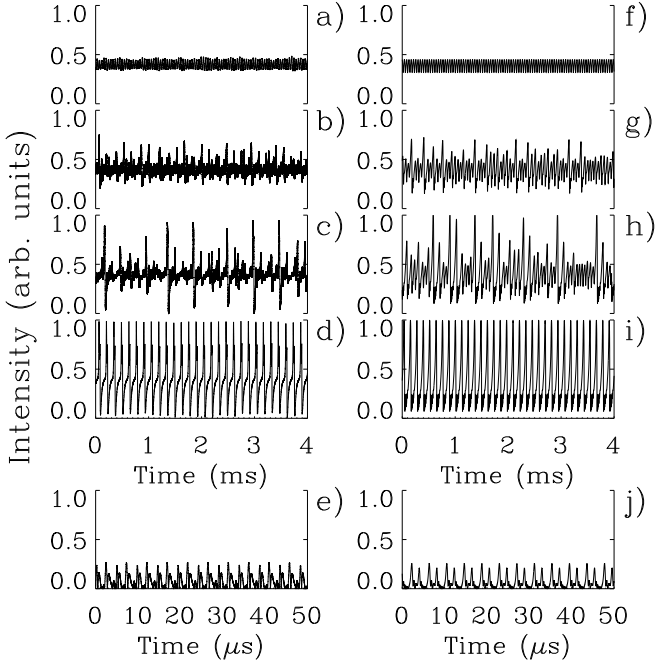}
\end{center}
\caption{Experimental time-series of the reflected intensity
for $P_{in}$=50 mW: a) $\Delta \nu =-3.9$MHz, b) $\Delta \nu =-3.6$MHz,
c) $\Delta \nu =-3.2$MHz, d) $\Delta \nu =-2.6$MHz, e) $\Delta \nu =0.8$MHz. Time-series of the reflected intensity $I = \vert \varepsilon \vert^2$
as obtained by numerical integration of Eqs. (\ref{eq1})-(\ref{eq3}), with $P_{in}$=50 mW and initial conditions ($I=1,
\phi=(\alpha/\omega^2) I, \theta=-\beta I$): f) $\delta_0=4$, g) $\delta_0=4.22$, h) $\delta_0=4.25$, i)
$\delta_0=4.29$, j)  $\delta_0$=14.9. Other parameters as in Fig. \ref{figu2}.}
\vspace{-.3cm}
\label{figu3}
\end{figure}

We now analyze in detail the dynamical regimes.
Figure \ref{figu3}a-e shows five traces of the reflected intensity as the detuning, $\vert \Delta \nu \vert$, between the cavity resonance and
the laser frequency is delicately decreased, approaching the resonance from the red side. The stationary intensity value loses
stability through a supercritical Hopf bifurcation and a small-amplitude quasi-harmonic limit cycle
is observed (\ref{figu3}a). In the vicinity of the bifurcation point, the characteristic frequency is around 20 kHz, i.e. much slower than
the radiation-pressure and cavity decay rates ($\sim$200 kHz). This indicates the existence of a further slow phenomenon playing a role in
the dynamics. The first Hopf bifurcation is then followed by a period-doubling cascade \cite{prenoi},
producing a sequence of small-amplitude periodic and chaotic attractors (see Fig. \ref{figu3}b).
As the cavity resonance is further approached a chaotic spiking regime
is observed, where the small amplitude chaotic oscillations are sporadically interrupted by large pulses
(\ref{figu2}c). The mean spikes rate increases as $\vert \Delta \nu \vert$ decreases until
a regime of large-amplitude regenerative oscillations is finally reached (Fig. \ref{figu2}d).
Note that the frequency of these oscillations is even slower than that of the Hopf quasi-harmonic limit cycle.
Moreover, the large pulses in the chaotic spiking regime have approximately the same amplitude and duration and roughly
resemble in shape those of the regenerative oscillations regime.
As we will later show, these qualitative features are compatible with the slow-fast dynamics described in Ref. \cite{noi}, where
each spike corresponds to a relaxation orbit in the phase-space and the chaotic spiking regime is the result of
chaotic canard explosions.

By further approaching the resonance condition, relaxation oscillations disappear and, at $\vert \Delta \nu \vert \approx 0$,
the system is stable. The narrow stability region around resonance separates two very different dynamical regimes,
both in amplitude and in frequency. On the red side of the resonance, we have the aforementioned slow-fast dynamics induced by the
interplay between radiation-pressure and photothermal effect. On the blue side, we find a periodic attractor,
roughly at the mechanical oscillator frequency, consisting of a train of decaying peaks (Fig. \ref{figu2}e).
This phenomenology has been previously observed in toroidal microcavities \cite{in6}
and it has been shown to be purely induced by the interplay between radiation-pressure and the intracavity optical field build-up
(at these frequencies, the photothermal contribution is indeed negligible).

The same sequence of regimes is found in the model (\ref{eq1})-(\ref{eq3}) for different values of $\delta_0$ .
The slow-fast dynamics is found by choosing initial conditions for the optical field close to its resonant value.
As in the experiment, the initial Hopf bifurcation takes place
when the mean normalized intracavity intensity is approximately equal
to $0.4$ (see Fig. \ref{figu3}f). Then, the system undergoes a period
doubling cascade leading to the birth of small-amplitude chaotic attractors (Fig. \ref{figu3}g).
Further increasing $\delta_0$, the mean amplitude of the attractors grows,
until that an erratic -- sensitive to initial conditions -- sequence of
pulses on top of a chaotic background takes place (see Fig. \ref{figu3}h).
For larger $\delta_0$, the mean firing rate increases until a periodic regime is finally reached (see Fig. \ref{figu3}i).
In Fig. \ref{figu3}j), we show the fast periodic dynamics with frequency close to that of the mechanical oscillator.

\begin{figure}
\begin{center}
\includegraphics*[width=1.0\columnwidth]{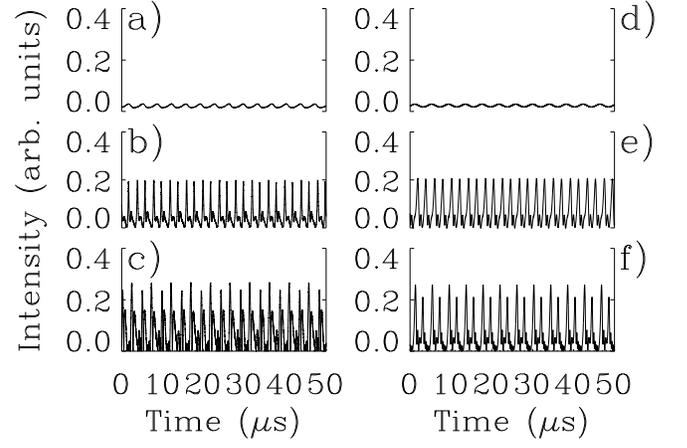}
\end{center}
\caption{Experimental time-series of the reflected intensity
for $P_{in}$=50 mW: a) $\Delta \nu =-1.1$MHz, b) $\Delta \nu =-3.4$MHz,
c) $\Delta \nu =-3.7$MHz. Time-series of the reflected intensity $I = \vert \varepsilon \vert^2$
as obtained by numerical integration of Eqs. (\ref{eq1})-(\ref{eq3}), with $P_{in}$=50 mW and initial conditions ($I=0.1,
\phi=(\alpha/\omega^2) I, \theta=-\beta I$): d) $\delta_0=6.5$, e) $\delta_0=4$, f)
$\delta_0=3$. Other parameters as in Fig. \ref{figu2}.}
%\vspace{-.3cm}
\label{figu4}
\end{figure}

When the detuning is varied in the opposite direction, approaching the cavity resonance from the blue side of the
spectrum, the slow-fast dynamics is never encountered. A new Hopf bifurcation produces
quasi-harmonic oscillations at the mechanical oscillator frequency (Fig. \ref{figu4}a), which evolve into the previously oberved train of decaying peaks
(Fig. \ref{figu4}b-c). These regimes coexist with the slow-fast dynamics, as confirmed also by numerical results (see Fig. \ref{figu4}d-f)
obtained by setting $\varepsilon=0$ as initial condition in the integration of Eqs. (\ref{eq1})-(\ref{eq3}).

\section{Bifurcation scenario}
\subsection{Coexisting attractors}
\begin{figure}
\begin{center}
\includegraphics*[width=1.0\columnwidth]{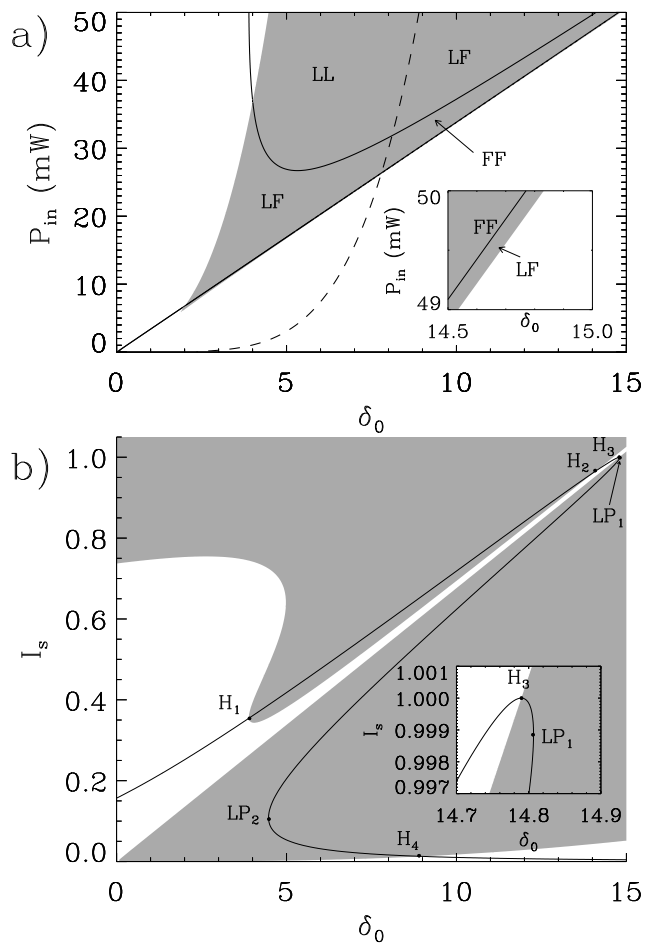}
\end{center}
\caption{a) Bistable regimes (shaded regions) calculated by means of Eq. (\ref{eq5}), together with the Hopf bifurcation boundaries (Eq. \ref{eq6}) of the
high intensity steady state (solid lines) and low-intensity stable state (dashed line), implicitly defined by Eq. (\ref{eq4}). $LL$: coexistence of non-stationary
solutions. $LF$: coexistence of a fixed point and a non-stationary solution. $FF$: coexistence of fixed points. b) Hopf instability domain (Eq. \ref{eq6})
and the stationary solutions (solid line) in the $(I_s,\delta_0)$ plane, for $P_{in}$=50 mW (other parameters as in Fig. \ref{figu2}). The intersections between the boundaries of the shaded region
and the $I_s$-curve determine the Hopf bifurcation points. In white regions the fixed points are stable.}
%\vspace{-.3cm}
\label{figu5}
\end{figure}

The steady state solutions of Eqs. (\ref{eq1})-(\ref{eq3}) are implicitly
defined by the cubic equation for the stationary optical intensity, $I_s$
\begin{equation}
I_s \left [1 + ( \delta_0 + \left (\frac{\alpha}{\omega^2} - \beta \right ) I_s)^2 \right ] - 1 = 0 \;
\label{eq4}
\end{equation}
Depending on the values of $\alpha$, $\beta$, $P_{in}$ and $\delta_{0}$, the
system can have either one or three fixed points. The change in the number
of stationary points occurs when
\begin{equation}
\left ( \frac{\delta_0^2 - 3}{9} \right )^3 = \left [ \frac{\alpha/\omega^2 - \beta}{2} + \frac{1}{3} \delta_0 \left( 1 + \frac{\delta_0^2}{9} \right) \right ]^2 \; .
\label{eq5}
\end{equation}
Eq. \ref{eq5} defines the boundaries of the region where the system has three stationary solutions
(shaded surface in Fig. \ref{figu5}a). These boundaries meet in a cusp point at $\delta_0 = \pm \sqrt{3}$ and
$(\alpha/\omega^2 - \beta) = \pm 8 / 3 \sqrt{3}$, where the latter define the maximum power for which the system
is monostable. At higher power values, the birth of the two new steady states can be due either to
to radiation pressure (when $(\alpha/\omega^2 - \beta) > 8 / 3 \sqrt{3}$) or to photothermal effect
(when $(\alpha/\omega^2 - \beta) < -8 / 3\sqrt{3}$), as in the present case.

In order to determine the nature of the coexisting attractors (fixed points and/or limit cycles)
we study the linear stability of the stationary states. The resulting characteristic equation for the eigenvalues $\Lambda$ is
\[
\Lambda^5 + a_1 \Lambda^4 + a_2 \Lambda^3 + a_3 \Lambda^2 + a_4 \Lambda + a_5 = 0 \; ,
\]
where the coefficients are given by
\begin{eqnarray}
a_1 &=& 2 + \omega_{PH} + \frac{\omega}{Q} \; , \nonumber \\
a_2 &=& \omega^2 + \omega_{PH} \left (2 + \frac{\omega}{Q} \right ) + 2\frac{\omega}{Q} +1 + \delta_{HOT}^2 \; ,\nonumber \\
a_3 &=& 2 \omega^2 + \omega_{PH} \left (\omega^2 + 2\frac{\omega}{Q} + 1 + \delta_{HOT}^2 \right ) + \; \nonumber \\
&+& \frac{\omega}{Q}(1 +\delta_{HOT}^2) - 2\beta\omega_{PH}\delta_{HOT} I_s \, \nonumber \\
a_4 &=& \omega_{PH} \left (2\omega^2 + \frac{\omega}{Q}(1 + \delta_{HOT}^2)\right ) + \; \nonumber \\
&+& \omega^2 \left ( (1 + \delta_{HOT}^2) + 2\frac{\alpha}{\omega^2} \left (1 -
\frac{\beta \omega_{PH}\omega}{\alpha Q} \right )\delta_{HOT} I_s \right )\, ,\nonumber \\
a_5 &=& \omega^2\omega_{PH} \left ( (1 + \delta_{HOT}^2) + 2\frac{\alpha}{\omega^2} \left (1 - \frac{\beta \omega^2}{\alpha} \right ) \delta_{HOT} I_s \right ) \; .
\nonumber \\
\end{eqnarray}
Here we have defined the "hot-cavity" detuning $\delta_{HOT}= \delta_0 + (\alpha/\omega^2 - \beta)I_s$ , i.e the detuning in the
presence of both radiation-pressure and photothermal effects. At a Hopf bifurcation the system must have a pair of purely imaginary eigenvalues
$\Lambda_{1,2} = \pm i \nu$. Demanding that the characteristic equation support these solutions establishes the following
expressions for the oscillation frequency $\nu$ and the Hopf bifurcation boundaries
%\begin{subequations}
\begin{equation}
\nu^2 = \frac{a_5 - a_1 a_4}{a_3 - a_1 a_2} \; \\
\label{eq6f}
\end{equation}
\begin{equation}
(a_2 a_1 - a_3) (a_3 a_4 - a_2 a_5) - (a_1 a_4 - a_5)^2 = 0 \, .
\label{eq6}
\end{equation}
A state diagram can be thus readily obtained by plotting in the
plane ($P_{in}$, $\delta_0$) the bistable region obtained by Eq.
(\ref{eq5}) together with the solutions of the algebraic system
defined by Eq. (\ref{eq4}) and Eq. (\ref{eq6}). Solid (dashed)
lines in Fig.~\ref{figu5}a) identify the Hopf bifurcation
boundaries for the high (low) intensity stable states defined by
Eq. (\ref{eq4}). The surface LL defines the region in which
non-stationary attractors coexist. Regions LF and FF refer to
fixed point/limit cycle bistability and fixed points bistability.
Notice that one of the branches of the Hopf bifurcation curve and
the boundary of the bistability region are nearly coincident, but
they never become tangent within the whole range of explored parameters
(see also the zoom reported in Fig. \ref{figu5}a). As a
consequence, we exclude the presence of a saddle-node-Hopf
bifurcation point. A more direct interpretation of the
experimental results can be obtained by plotting the Hopf
bifurcation boundaries in the plane ($I_s$, $\delta_0$) together
with the steady states solutions, $I_s$, keeping fixed the
injected power. Approaching the resonance from the red side (from
left to right in Fig.~\ref{figu5}b), the first crossing of the
$I_s$-curve with the boundary defined by Eq. (\ref{eq6}), will
drive the system unstable through a supercritical Hopf bifurcation
$H_1$, where a finite-frequency limit cycle starts to grow.
Further increasing $\delta_0$ leads to the inverse bifurcation
$H_2$ and the system passes from the oscillatory dynamics to a new
steady state solution. Chaotic canard explosions and the
subsequent relaxation oscillation regime are encountered between
$H_1$ and $H_2$. The system remains stable until the resonant
condition, $I_s = 1$, is reached. However, just beyond resonance,
an Hopf bifurcation occurs ($H_3$) and a different oscillatory
regime takes place, which coexists with a stable fixed point (the
low intensity solution). As discussed before, the bifurcation
$H_3$ occurs before --though very close to --, the limit point
$LP_1$, where the system passes from three to one steady state
solution (see inset in Fig. \ref{figu5}b). The steady states
between $H_3$ and $LP_1$, have one real and two pairs of complex
conjugate eigenvalues (one with positive and the other with
negative real part). The real eigenvalue, which is negative for
equilibria on the upper branch of the steady-state curve and
positive for those on the middle one, vanishes at the limit point
$LP_1$. Therefore, at $LP_1$ the coalescence and disappearance of
two saddle-foci takes place. Similar arguments can be used for describing the behaviour in the vicinity of the lower limit point $LP_2$.

On the other hand, by approaching the resonance from the blue
side, we follow the lower stable branch of the $I_s$ curve up to
the supercritical Hopf bifurcation $H_4$. For $\delta_0$ values
beyond this bifurcation, we have the coexistence of limit cycles
born in $H_1$ and $H_4$. By means of Eq. (\ref{eq6f}), we can
evaluate the limit cycle frequencies at the bifurcations points
$H_1$ and $H_4$, and we find $\nu_1 \sim 0.1 \approx 22$ kHz and
$\nu_4 \sim 1.1 \approx$ 240 kHz in good agreement with the
experimental results.

\subsection{Slow-manifold and chaotic canards}

\begin{figure}
\begin{center}
\includegraphics*[width=1.0\columnwidth]{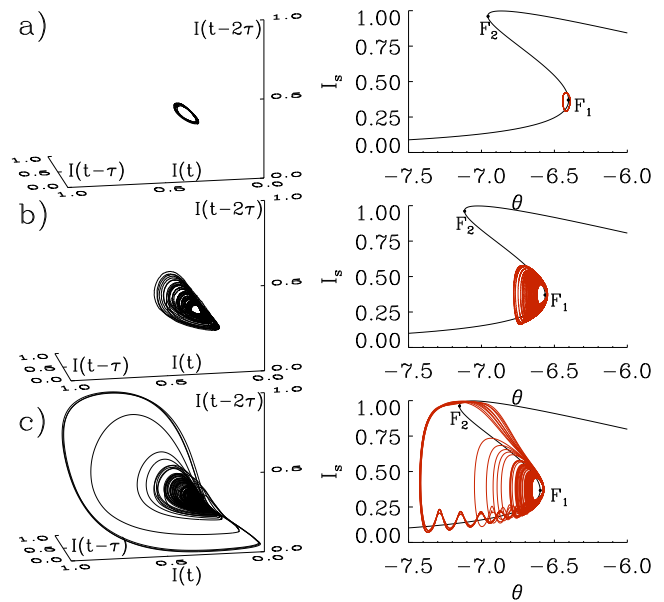}
\end{center}
\caption{Left panels: three-dimensional reconstruction of the phase-portrait through the Ruelle-Takens embedding technique with time delay
$\tau$=1.74 $\mu$s, from experimental time-series of the optical intensity. a) Hopf quasi-harmonic cycle (see Fig. \ref{figu3}a), b) small-amplitude chaotic
attractor (see Fig. \ref{figu3}b)
and chaotic spiking regime (see Fig. \ref{figu3}c). Right panels: the corresponding numerical phase-space trajectories (light curves) together with the slow manifold (black
curve) (\ref{eq4}) in the $(I, \theta)$ plane. Parameters as in Fig. \ref{figu3}f-h.}
%\vspace{-.3cm}
\label{figu6}
\end{figure}

The linear stability analysis provides information about
the dynamic behavior only within a small range around the
bifurcation point, where the system exhibits a small amplitude harmonic limit cycle. Well beyond the Hopf bifurcation, a
chaotic spiking regime is observed, consisting of erratic relaxation spikes interrupting periods of irregular small-amplitude quasi-harmonic oscillations.
The blow-up of such a regime can be understood by analyzing the limit $\omega_{PH} \simeq 0$ in
Eqs. (\ref{eq1})-(\ref{eq3}). All the parameters
appearing in the equations are $O$(1) quantities except for the
nomalized photothermal rate ($\omega_{PH}$) and the mechanical reaction rate ($\omega/Q$)
which are of the order of $10^{-4}$. In these conditions, the system is described by a slow-fast
dynamical system and geometric theory of singular perturbation is readily applicable. Since
the evolution of $\theta$ is slow with respect to that of the other variables, we can analyze
separately slow and fast motions. The fast motion follows Eqs. (\ref{eq1})-(\ref{eq2}) (fast
subsystem), where $\theta$ is constant and has to be considered as a bifurcation
parameter. The fixed points of this dynamical
subsystem lay on the one-dimensional manifold, $\Sigma$, implicitly defined by Eq. (\ref{eq4}), and
it is on this manifold where the slow dynamics described by Eq. (3) take place.
By linearizing the fast subsystem evolution equations around points of the set $\Sigma$,
it is simple to see that the equilibria of the fast subsystem are stable if
\begin{equation}
f_{\theta}(I_s) \equiv \frac{2 \alpha}{\omega^2} (\delta_0 + \frac{\alpha}{\omega^2} I_s + \theta) > -1
\end{equation}
and unstable if $f_{\theta}(I_s) < -1$. The boundaries separating the stable branches of the slow
manifold from the unstable part (the turning points $F_{1,2}$ in Fig. \ref{figu6}) are
determined by Eq. (\ref{eq4}) together with the condition $f_{\theta}(I_s) = -1$.
Therefore, the slow manifold is composed of two attracting branches, separated by a
repelling branch. On the basis of this analysis, we can qualitatively understand the onset of the
chaotic spiking regime in our system. In Fig. \ref{figu6} we plot the reconstructed attractors
from experimental time series (left panels) and the corresponding numerical phase-space trajectories
together with the slow manifold $\Sigma$ (right panels).
The initial Hopf bifurcation (Fig. \ref{figu6}a) is followed by a period-doubling cascade
producing a sequence of small-amplitude periodic and chaotic
attractors (see Fig. \ref{figu6}b), that develops before relaxation oscillations arise.
As $\vert \Delta \nu \vert$ ($\delta_0$) is decreased (increased), the mean amplitude of the attractors grows, until the chaotic fluctuations are sufficiently
large to eventually "trigger" the fast dynamics. As a result, chaotic canard
explosions are observed, i.e. an erratic -- sensitive to initial
conditions -- sequence of canard orbits separated by an irregular
number of small amplitude quasi-harmonic oscillations.
The typical features of canards are evidenced in Fig. \ref{figu6}c). The trajectories crossing the fold
point $F_1$ do not jump immediatly to the other attractive branch of $\Sigma$, but continue moving on the slow time scale along the
unstable portion of the slow manifold. As the detuning is further decreased, the amount of time spent by
the system in the vicinity of the repelling part of the manifold diminishes and the orbits approach the relaxation
oscillation full cycle.

\section{Conclusions}
We have presented a complete characterization of the multiple-time scale dynamics
induced by radiation pressure and photothermal effects in suspended mirror resonators.
We have completed the analysis of the slow-fast dynamics, first observed in Ref. \cite{prenoi},
showing that chaotic spiking sequences are the result of canard explosions in a higher-dimensional phase space,
in which trajectories are rapidly attracted towards the stable branches of a one-dimensional {\bf S}-shaped manifold.
The dynamics here described, as well as the structure and the stability of the slow manifold,
are fully compatible with those numerically observed in a 3D slow-fast system governed by one slow and two fast variables \cite{noi}.
Moreover, we have shown that this regime coexists with other non-trivial periodic solutions, whose characteristic time-scales
are substantially faster. In particular, the steady resonant condition, i.e. the operation region required
in quantum optics experiments, coexists with a limit cycle solution.
Therefore, our results find immediate applications in the field of optomechanical experiments,
in order to design control schemes able to mantain the system close enough to the resonant steady state.
The recent progresses in the photonic crystal engineering have opened the prospect to build arrays of many coupled optomechanical
oscillators integrated on a chip. Even in the classical regime, one could thus be able to explore the collective
dynamics of arrays consisting of many coupled optomechanical cells. In the presence of field dynamics and radiation pressure
it has been theoretically shown that optomechanical arrays can display synchronization,
and that they can be described by an effective Kuramoto-type model \cite{prl107}.
In this framework, the dynamics we observe is even more interesting, since it involves the generation of irregular
(deterministic) spike sequences, which are known to play an important role in neural information processing. Therefore, the possibility to study synchronization phenomena
between chaotically spiking units could be of interest in different fields ranging from physics to dynamical system theory
and neuroscience.

\end{document}